\documentclass{osa-article}

\usepackage{graphicx}
\usepackage{caption}
\usepackage{subcaption}
\usepackage{float}
\usepackage{dirtytalk}

\usepackage{mathrsfs}
\usepackage{gensymb}
\usepackage{lineno}
\usepackage[labelsep=period]{caption}

\journal{oe}

\begin{document}
\title{Effect of imperfect homodyne visibility on multi-spatial-mode two-mode squeezing measurements}

\author{Prasoon Gupta,\authormark{1} Rory W. Speirs,\authormark{1} Kevin M. Jones,\authormark{2} and Paul D. Lett\authormark{1,3,*}}

\address{
\authormark{1}Joint Quantum Institute, National Institute of Standards and Technology and the University of Maryland, College Park, Maryland 20742, USA\\
\authormark{2}Department of Physics, Williams College, Williamstown, Massachusetts 01267, USA\\
\authormark{3}Quantum Measurement Division, National Institute of Standards and Technology, Gaithersburg, Maryland 20899, USA}

\email{\authormark{*}paul.lett@nist.gov} 



\begin{abstract}
We study the effect of homodyne detector visibility on the measurement of quadrature squeezing for a spatially multi-mode source of two-mode squeezed light. Sources like optical parametric oscillators (OPO) typically produce squeezing in a single spatial mode because the nonlinear medium is within a mode-selective optical cavity. For such a source, imperfect interference visibility in the homodyne detector couples in additional vacuum noise, which can be accounted for by introducing an equivalent loss term.
In a free-space multi-spatial-mode system imperfect homodyne detector visibility can couple in uncorrelated squeezed modes, and hence can cause faster degradation of the measured squeezing. We show experimentally the dependence of the measured squeezing level on the visibility of homodyne detectors used to probe two-mode squeezed states produced by a free space four-wave mixing process in $^{85}$Rb vapor, and also demonstrate that a simple theoretical model agrees closely with the experimental data. 
\end{abstract}

\section{Introduction}
Quadrature squeezed states of light are useful resources for continuous variable quantum information protocols \cite{Furusawa706}, precision metrology, and gravitational wave detection \cite{PhysRevLett.116.061102, PhysRevLett.59.278}. The measured squeezing level places limits on the fidelity of quantum information protocols \cite{Furusawa706, PhysRevA.64.022321}, and the sensitivity of phase measurements \cite{PhysRevA.86.023844, PhysRevLett.59.278, Anderson:17, Gupta:18}. It is therefore important to understand the factors that limit the measurement of the squeezing level.

Homodyne detection involves the overlap of a strong local oscillator (LO) with a signal beam whose quadrature we wish to measure. The visibility of the optical interference in a homodyne detector is determined by how the intensity distributions of the signal beam and the LO overlap and how well their phase fronts match.  A visibility of unity corresponds to perfect mode matching, and a lower visibility signifies some degree of mismatch between the modes. A mismatch in the modes of the LO and the signal beam causes partial overlap of the LO with modes orthogonal to the signal beam spatial mode, making the homodyne detector sensitive to the noise in those orthogonal modes.

A squeezed light system such as an optical parametric oscillator (OPO) typically produces squeezing in a single spatial mode due to the presence of a mode-selective optical cavity. In a single mode source, any mode mismatch between the signal beam and the LO causes partial overlap of the LO with vacuum modes. In this case, the effect of imperfect visibility on the measurement of quadrature squeezing can be modeled by an equivalent optical loss term that depends on  $\mathcal{V}^2$, where $\mathcal{V}$ is the visibility of the homodyne detector \cite{PhysRevA.67.033802,Aoki:06, Polzik1992, Yonezawa1514}. Squeezed light sources like four-wave mixing (4WM) in Rb vapor have high single-pass gain and do not generally use a cavity, and hence they produce multi-spatial-mode squeezed states of light. In such a system, the signal beam modes (normally chosen with the help of a seed beam) are surrounded by other squeezed modes. If the spatial modes of the signal beam and the LO do not match, then the partial overlap of the LO with the modes orthogonal to the signal beam couples in noise from these orthogonal modes which can be noisier than vacuum modes.  

Previous theoretical work on the effects of mode mismatch between the signal and the LO in homodyne detection \cite{LaPorta, Bennink, Wasilewski} has dealt primarily with the pulsed generation of single-mode squeezing and the mixing-in of extra temporal modes as the nonlinear process used to generate squeezing produces a pulse that is shaped differently from the pump. That is, a one-dimensional (time) case is treated, where the mode shape of the signal can change, for example with pump power, and hence the overlap with the LO changes as well.  This can lead to a measurement of squeezing that is reduced at higher powers.  This work provides an analogy to what we see here considering the LO overlap with selected spatial modes in the case of two-mode squeezing.  If we use the LO on one beam to designate the spatial modes of the signal, the mismatch of the LO for the twin beam from selecting exactly the conjugate modes can be considered analogous to the local oscillator pulse overlapping neighboring squeezed temporal modes in the case of pulsed single-mode squeezing.  In each case the local oscillator then measures excess noise in squeezed but uncorrelated neighboring modes.  Instead of having the problem of using a Gaussian LO to measure non-Gaussian modes at high gain as is typically discussed, however, our problem is not in having the \say{wrong shape} for the LO, but that there are many relatively small strongly-squeezed modes, and it is a \say{placement error} of the second LO in selecting the desired modes.  The selection of the desired modes does not change as the pumping is increased in the present work, but the noise penalty for a fixed small mode mismatch does.  We compare the predictions of such a theoretical model to measurements in a two-mode squeezing system.

In the case of two-mode squeezing, we measure the squeezing in the joint quadratures of two correlated modes, by using two homodyne detectors, one on the probe and one on the conjugate, and electrically combine the signals. Individually, the probe and conjugate are displaced thermal beams which have higher quadrature noises than vacuum. In a multi-spatial-mode system, the probe and the conjugate beams are surrounded by spatial modes that show squeezing in their respective joint quadratures. The local oscillators for each of the probe and conjugate beams need to address the correlated spatial modes of the probe and the conjugate, as well as to address the same relative phase in each of them.  Any mismatch in the spatial modes of the two LOs and the probe and the conjugate would couple in the quadrature noises of the spatial modes orthogonal to the correlated probe and conjugate spatial modes. These additional spatial modes in the probe and conjugate homodyne detectors could be uncorrelated with each other, and hence add excess noise to the joint quadrature noise of the correlated probe and conjugate beams. These additional modes could also be correlated but out of phase with the signal modes, and hence can add excess noise to two-mode squeezing. In this work, we show how the measurement of squeezing in a multi-spatial-mode two-mode squeezed light system depends on the visibilities of the homodyne detectors.

\begin{figure}
	\centering
	\includegraphics[width=0.75\linewidth]{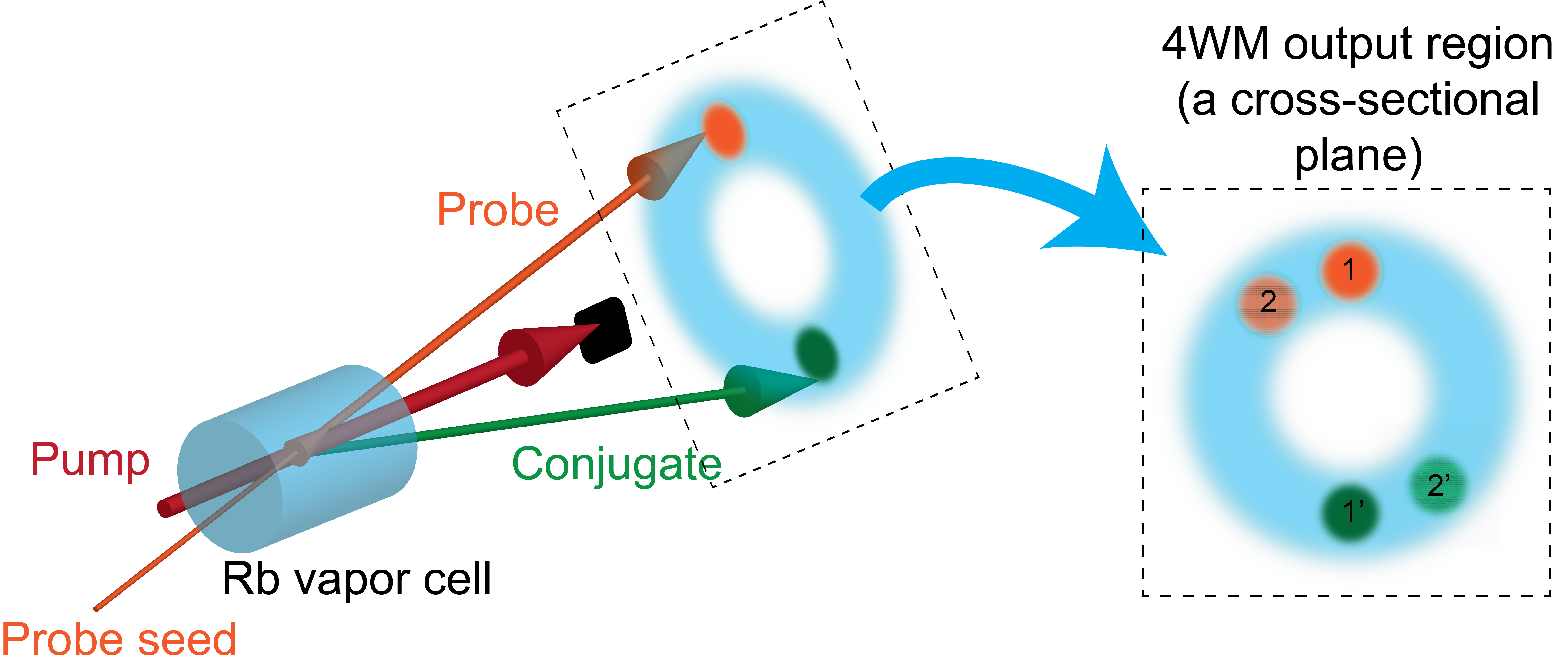}
	\caption{Schematic of the 4WM process. The blue annulus represents the output region where there is significant gain. Pairs 1-1' and 2-2' are quantum correlated, while there is no quantum correlation between the modes 1-2' and 2-1'.}
	\label{fig:modesIn4WM}
\end{figure}

\section{Experiment}
We use a double lambda 4WM process in \textsuperscript{85}Rb vapor to produce non-degenerate probe and conjugate beams which are in a two-mode squeezed state\cite{McCormick:07, Boyer544}. We send a strong pump beam ($\approx\!500$ mW) through hot Rb vapor along with a probe seed beam (derived from the pump and frequency-shifted by the ~3 GHz ground state splitting in $^{85}$Rb with an acousto-optic modulator) that intersects with the pump at an angle of $\approx\!1\degree$. The 4WM process amplifies the probe beam and produces a conjugate beam, as shown in Fig.~\ref{fig:modesIn4WM}. The 4WM process uses the $\chi^{(3)}$ nonlinearity in Rb vapor to annihilate two pump photons and produce a probe and a conjugate photon. The probe and conjugate photons are produced simultaneously and are correlated, which forms the basis for two-mode squeezing in the system. The angle between the probe seed and the pump can be varied slightly and the probe will still experience appreciable gain. Additionally, the gain will remain unchanged if the probe seed is rotated around the pump, so long as the angle between them remains the same. This defines the annular output region, shown in blue in Fig.~\ref{fig:modesIn4WM}. In the absence of a bright probe seed, the modes in this output region are seeded by vacuum but still experience gain, and thus generally have noise levels greater than vacuum. The probe spot produced by a bright seed beam is quantum correlated with a bright conjugate spot on the opposite side of the annulus but not with other spots on the annulus.

\begin{figure}
	\centering
	\begin{subfigure}[b]{0.4\textwidth}
		\centering
		\includegraphics[width=\linewidth]{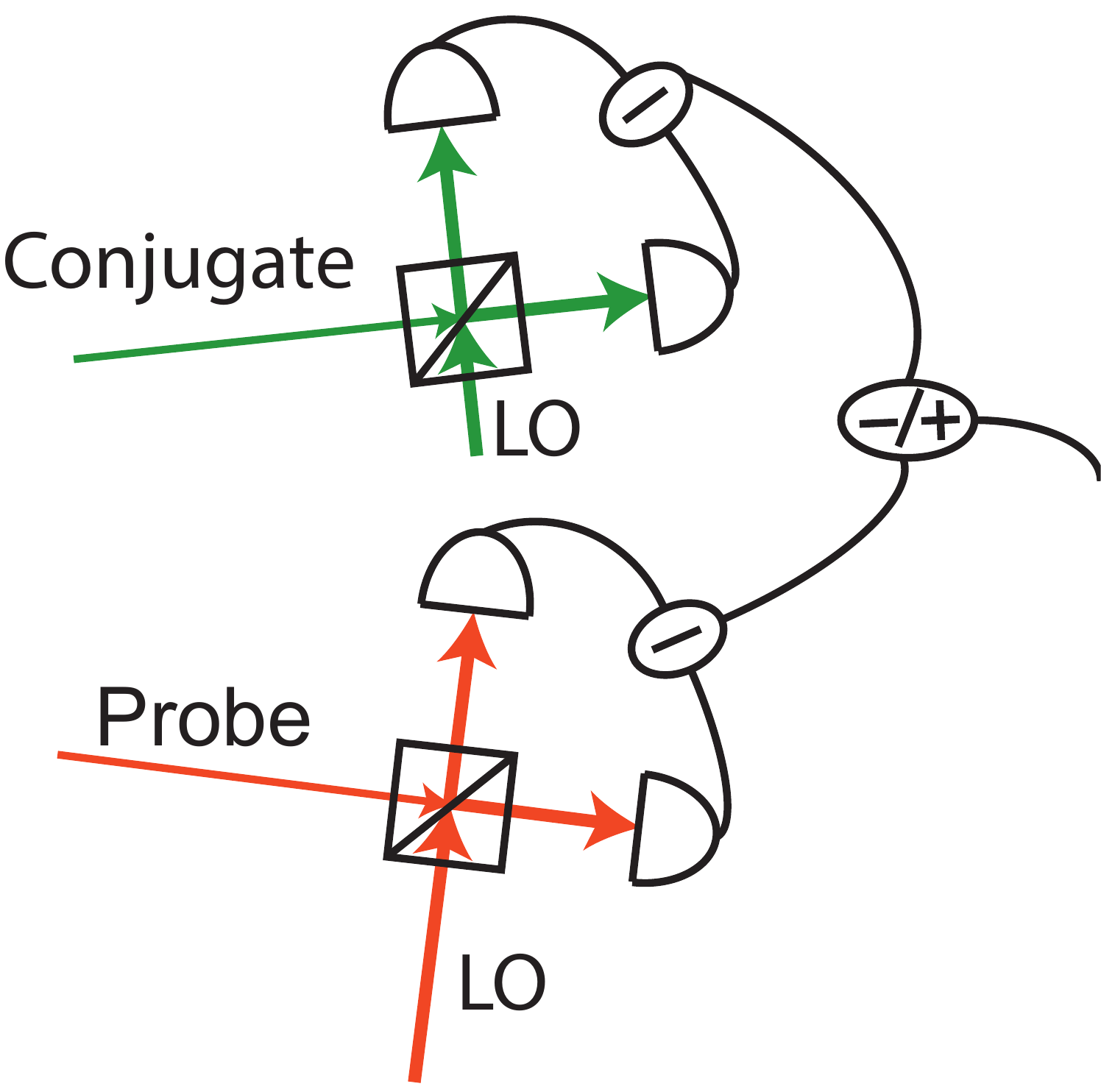}
		\caption{}
		\label{fig:jointHomodyneDetectionSetup}
	\end{subfigure}
	\qquad
	\qquad
	\centering
	\begin{subfigure}[b]{0.4\textwidth}
		\centering
		\includegraphics[width=\linewidth]{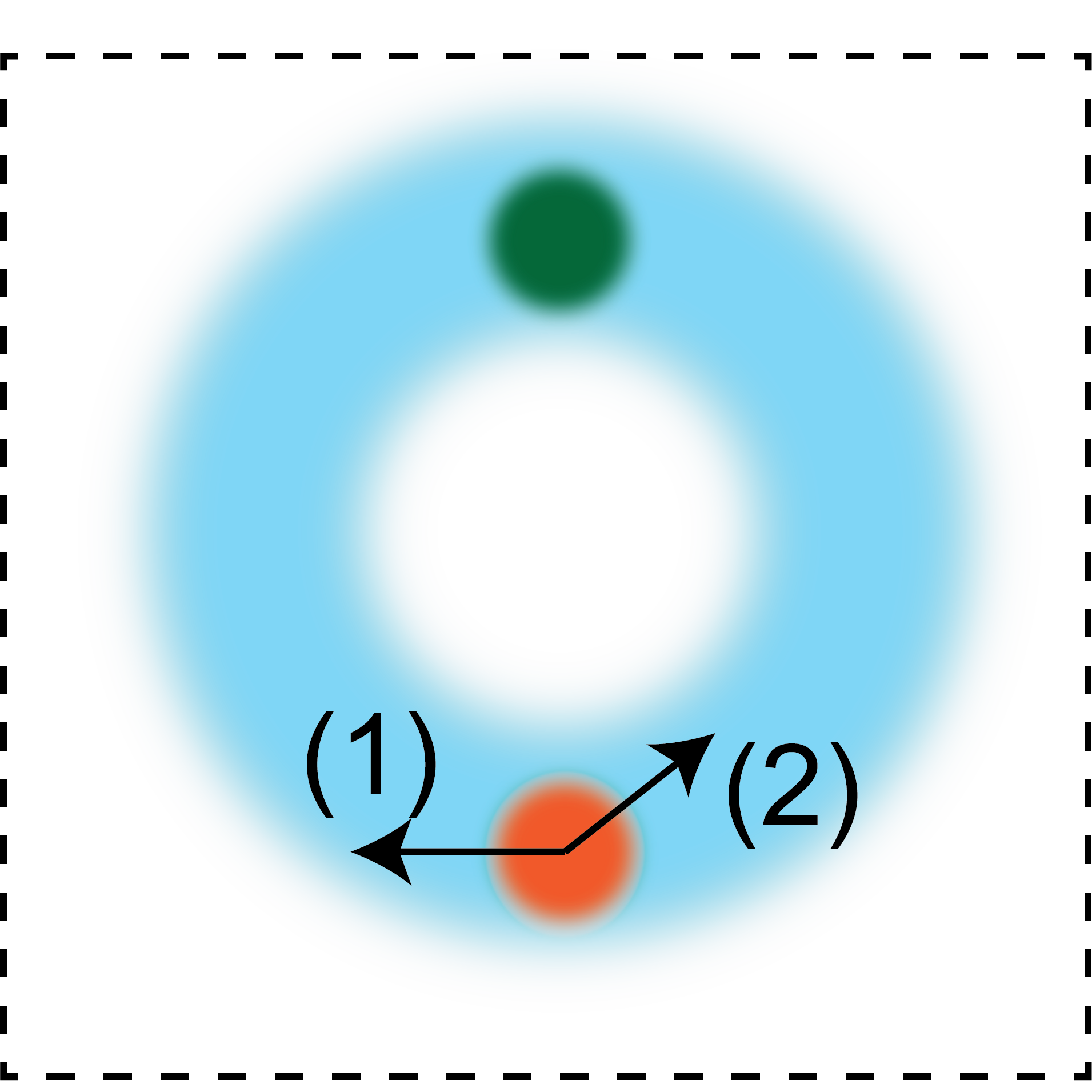}
		\caption{}
		\label{fig:beamDisplacement}
	\end{subfigure}
	\caption{(a) A setup for joint homodyne detection on the probe and the conjugate beams. (b) Displacement of probe LO beam (orange region) along two different directions in the cross-sectional plane using a three-axis piezo-electric mirror. The conjugate LO (green region) is kept at a constant position.}
\end{figure}

To measure the joint quadrature of the two beams, we use two homodyne detectors, one on the probe, one on the conjugate, and electrically combine the signals (Fig.~\ref{fig:jointHomodyneDetectionSetup}). We measure the noise at a frequency of 1 MHz with a resolution bandwidth of 100 kHz using a radio-frequency spectrum analyzer. We generate the probe and conjugate local oscillators (LO) using a separate 4WM process in the \textsuperscript{85}Rb vapor. The two LO beams are overlapped with the corresponding probe and the conjugate signal beams in the two homodyne detectors (Fig.~\ref{fig:jointHomodyneDetectionSetup}). Generating LOs in a separate 4WM process similar to the one used for generating the signal beams (the probe and the conjugate) helps match the LO and signal spatial profiles and phase fronts in the homodyne detectors, which improves the interference visibility. This becomes particularly important in the probe homodyne detector as the probe beam suffers lensing due to the cross-Kerr modulation induced by the pump beam\cite{Boyer544}. We keep the power of each of the LOs at approximately 500$~\mu$W and that of the probe and the conjugate in the signal pair at less than 1$~\mu$W. We control the powers of these beams by controlling the power of the seed beam and the gain in each 4WM process.
While the LOs generated in this way are displaced thermal states and have excess noise above that of a coherent state, the displacement is large with respect to the intensity of the squeezed vacuum and balanced homodyne measurements on a detector with good common mode rejection takes out this excess classical noise to a large degree, as can be seen in Ref.\cite{Boyer544}. Coherent-state local oscillators could, alternatively, be generated by frequency-shifting part of the pump beam in the same way that the probe seed is generated, but then the distortions due to the self-focusing of the signal beams in the Rb cell would affect the visibilities unless this was compensated for. The difference due to the excess noise from using the displaced thermal beam LOs would only be noticeable in our case when the measured squeezing is more than -10\,dB. 

To study the effect of homodyne detector visibility on the measured level of squeezing, we keep the visibility at the conjugate homodyne detector fixed, and change the visibility in the probe homodyne detector only. To adjust the visibility, we use a 3-axis piezo-electric mirror in the probe LO beam path to displace the LO beam along the two directions in a plane as shown in Fig.~\ref{fig:beamDisplacement}. In the homodyne detectors, we measure the visibility by making the LO and the signal beam equal in power and then measuring the output of one of the ports of the beam splitters. We measure the maximum power and minimum power coming out of the port as we change the phase difference between the two beams in the homodyne detector setup. To calculate the visibility, we use the expression:
\begin{equation}
\mathcal{V}=\frac{\textup{(max power-min power)}}{\textup{(max power+min power)}}.
\end{equation}

\section{A model for squeezing measurement in a multi-spatial-mode system}

Before describing how the homodyne detector visibility affects the measured two-mode squeezing level, we analyze its effect on the measurement of noise on a single beam. Here we choose the probe beam for the study (the conjugate would have served equally well). The probe is a displaced thermal beam, which is noisier than vacuum. The excess noise over that of the vacuum depends on the gain of the 4WM process and the optical loss in the system. The probe quadrature noise is independent of the measured quadrature, and hence does not depend on the phase difference between the probe and its LO in the homodyne detector. While the intrinsic noise on the beam is set by the optical gain and loss in the source, the measured noise is also affected by the visibility in the probe homodyne detector. 

If the probe beam were surrounded by vacuum modes - a characteristic of a spatially single-mode system - any mismatch between the LO and the probe spatial mode would couple vacuum noise into the homodyne detector. The overlap with the quieter vacuum modes would result in a reduction of the measured noise as the visibility of the homodyne detector is reduced.

Our 4WM system generates light in multiple spatial modes (Fig.~\ref{fig:modesIn4WM}), so the region surrounding the bright displaced thermal probe beam is populated with vacuum-seeded thermal light, which is uncorrelated with the desired conjugate mode. Hence in the homodyne detector a slight misalignment of the local oscillator will reduce its overlap with the bright seeded probe mode and will couple in vacuum-seeded modes, which are uncorrelated with the desired mode. These vacuum-seeded modes in the vicinity of the probe mode typically have experienced approximately the same gain as the probe mode, and thus have approximately the same noise level as the probe mode. Therefore, when the LO is shifted away from the bright probe beam the measured noise level should remain approximately constant even though the detector visibility has been decreased.

\begin{figure}[H]
	\centering
	\begin{subfigure}{\textwidth}
		\centering
		\begin{subfigure}{0.4\textwidth}
			\centering
			\text{Direction (1)}
			\includegraphics[width=\linewidth]{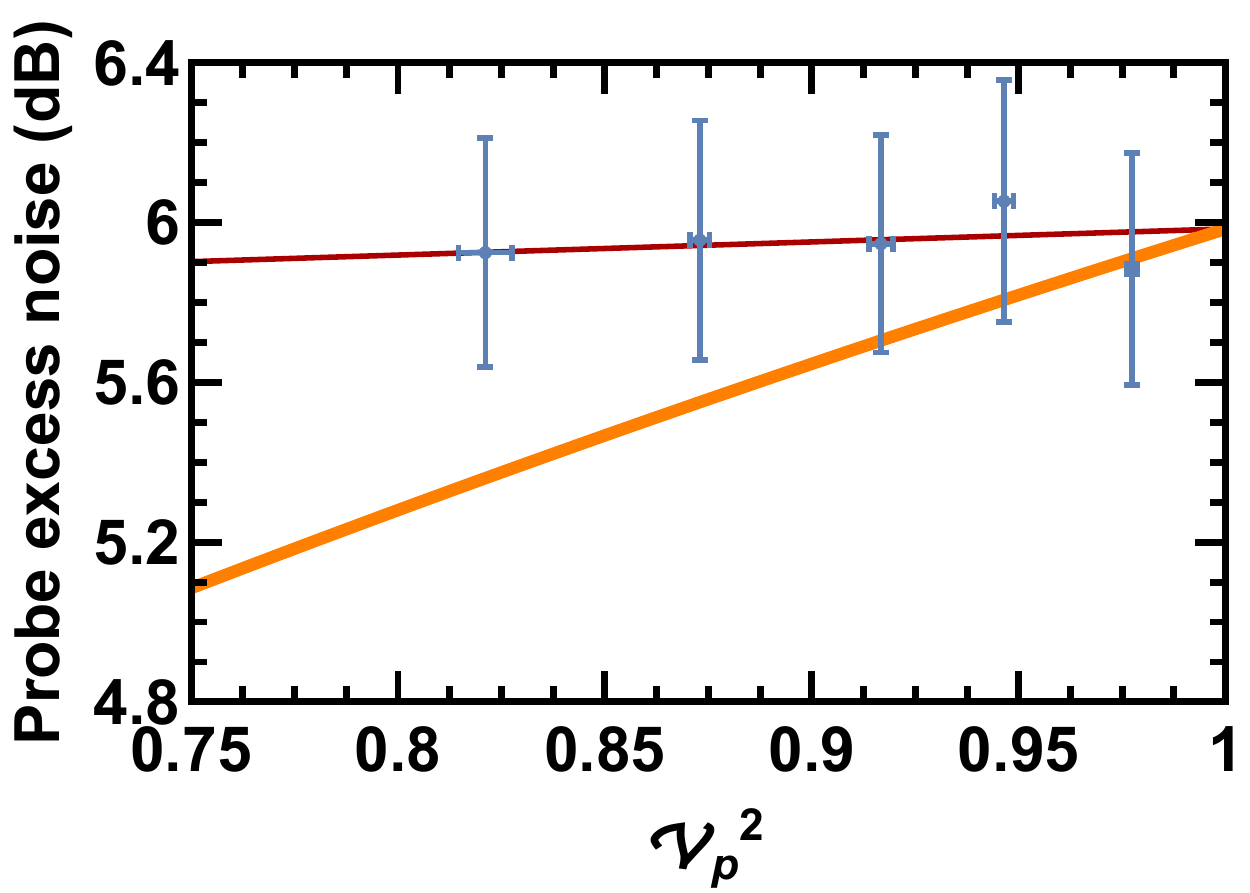}
			\caption{}
			\label{fig:single_beam_noise_dir1}
		\end{subfigure}
		\hfill
		\begin{subfigure}{0.4\textwidth}
			\centering
			\text{Direction (2)}
			\includegraphics[width=\linewidth]{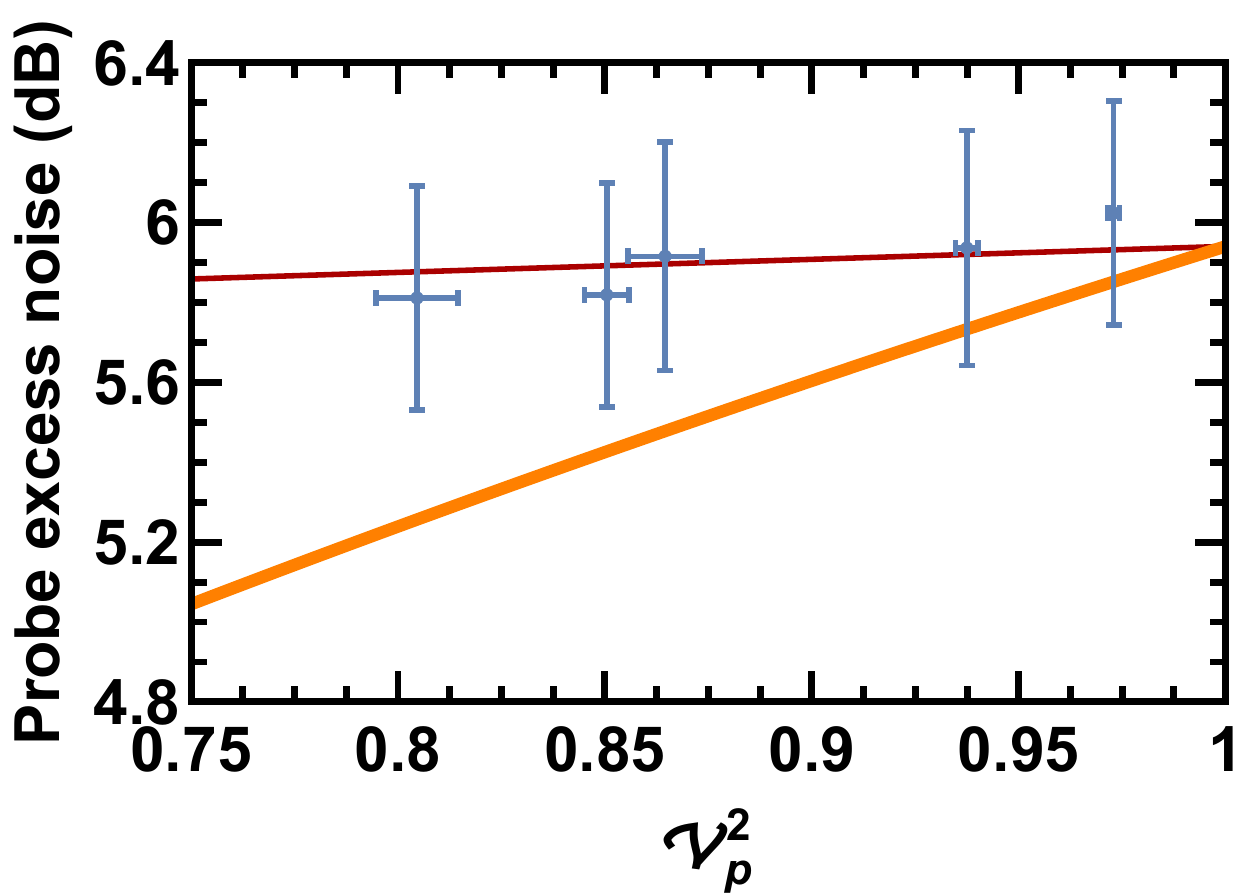}
			\caption{}
			\label{fig:single_beam_noise_dir2}
		\end{subfigure}
	\end{subfigure}
	\centering
	\begin{subfigure}{\textwidth}
		\centering
		\begin{subfigure}{0.4\textwidth}
			\centering
			\includegraphics[width=\linewidth]{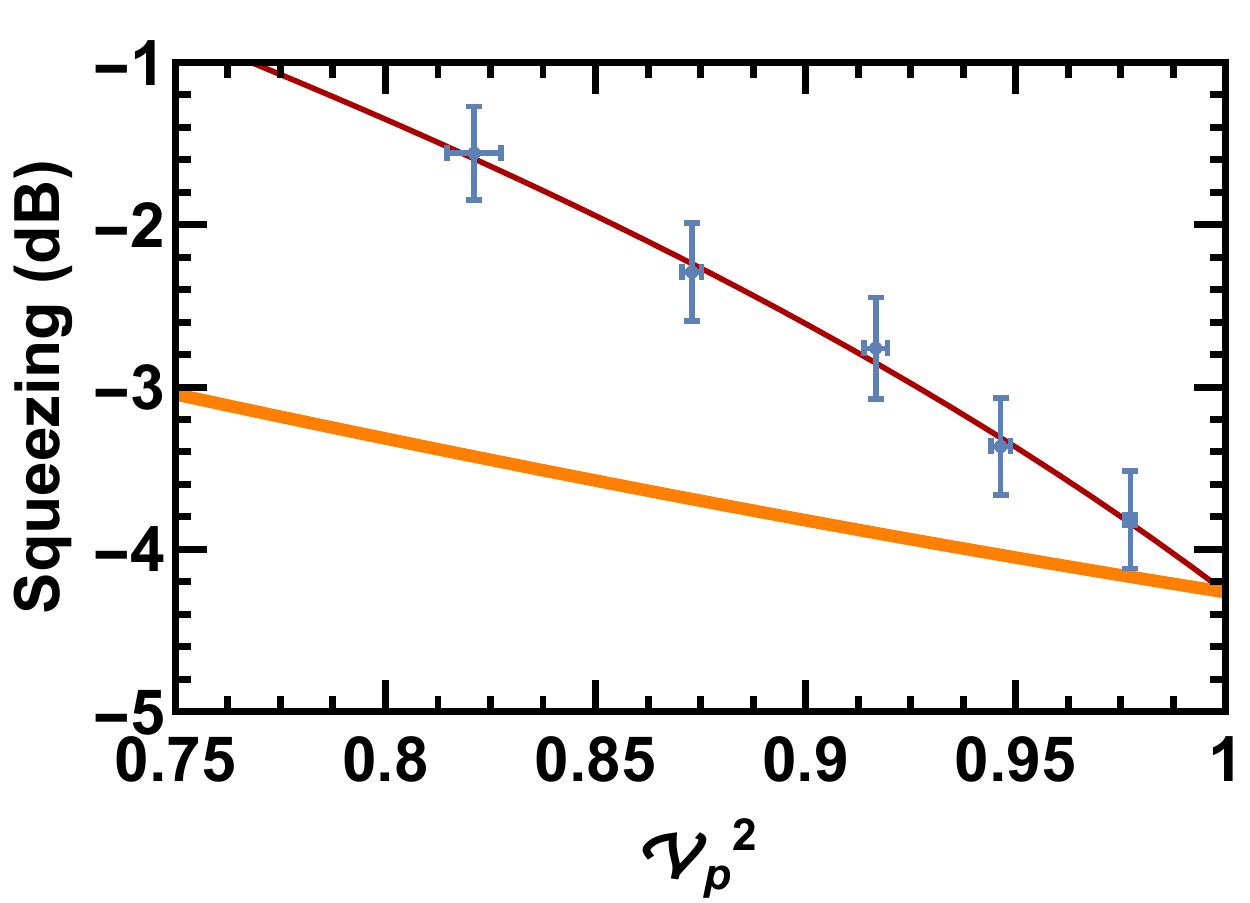}
			\label{fig:squeezingCh1}
			\caption{}
			\label{fig:squeezingWithVisibility_dir1}
		\end{subfigure}
		\hfill
		\begin{subfigure}{0.4\textwidth}
			\centering
			\includegraphics[width=\linewidth]{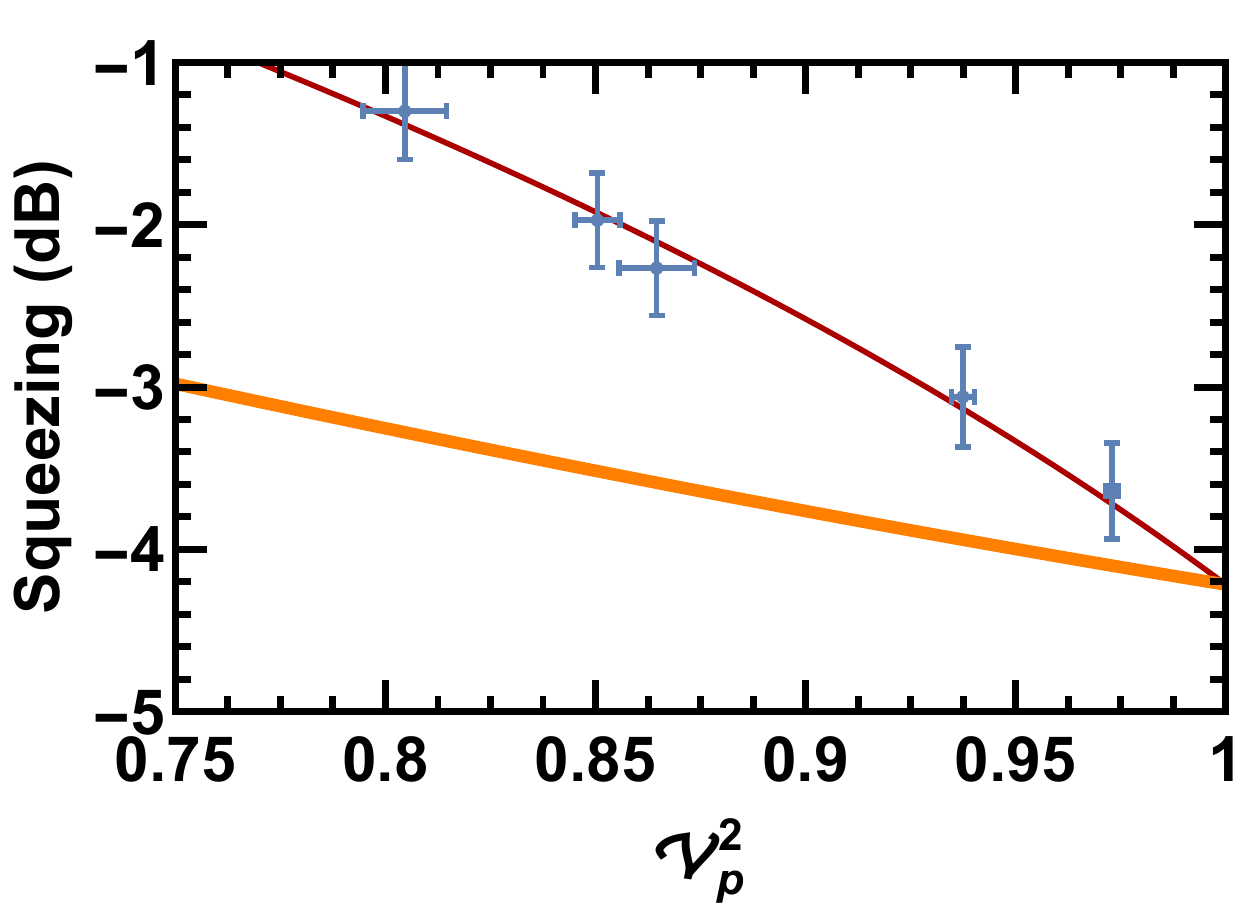}
			\label{fig:squeezingCh2}
			\caption{}
			\label{fig:squeezingWithVisibility_dir2}
		\end{subfigure}
	\end{subfigure}
	
	\begin{subfigure}{\textwidth}
		\centering
		\begin{subfigure}{0.4\textwidth}
			\includegraphics[width=\linewidth]{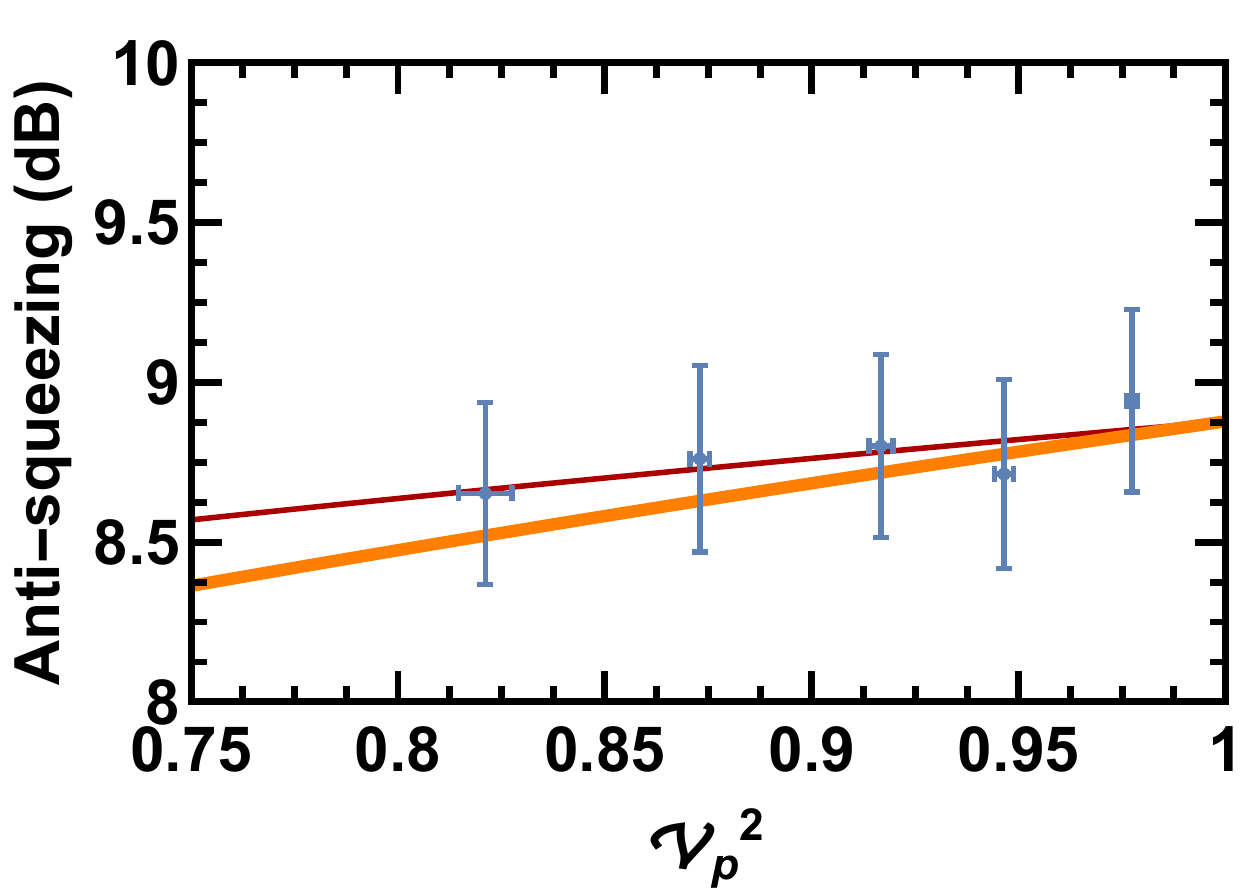}
			\caption{}
			\label{fig:antiSqueezingWithVisibility_dir1}
		\end{subfigure}
		\hfill
		\begin{subfigure}{0.4\textwidth}
			\includegraphics[width=\linewidth]{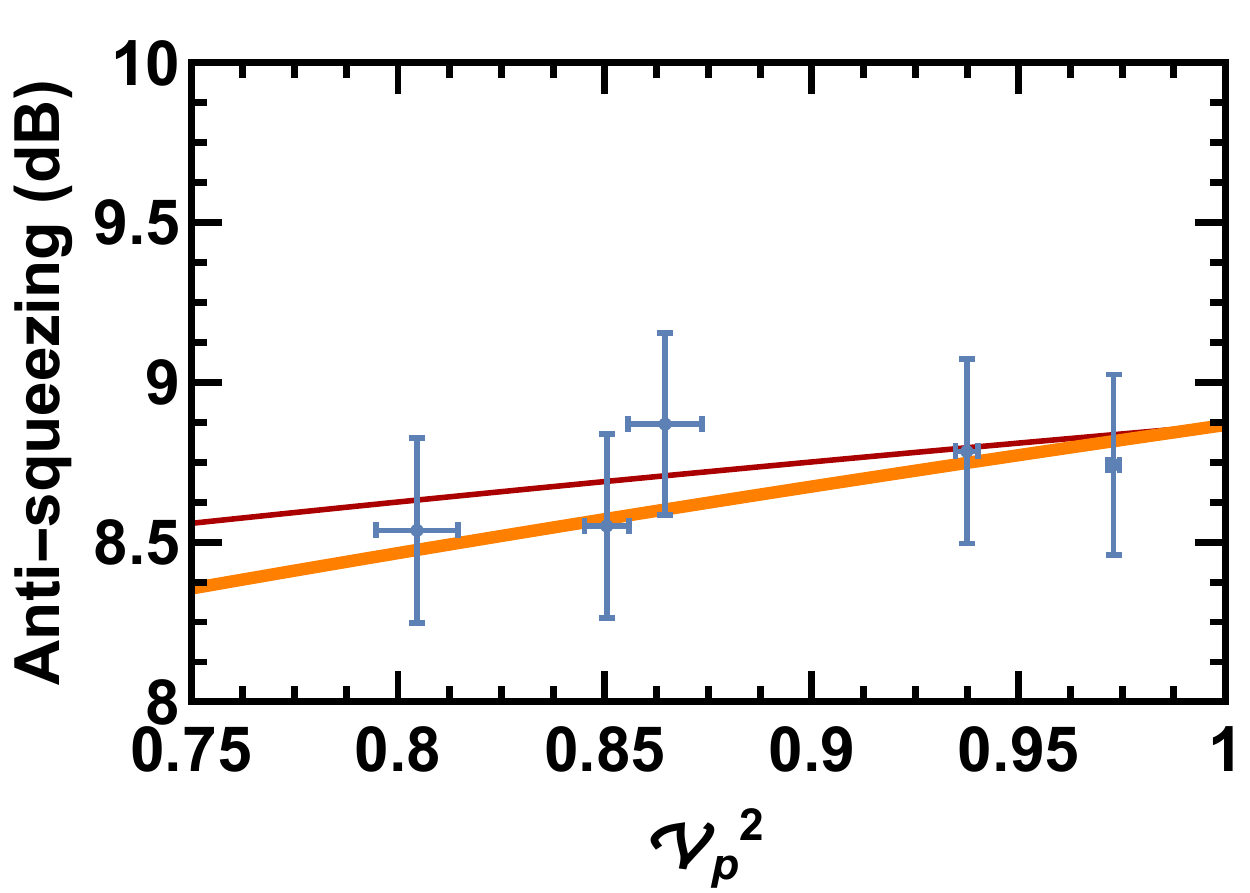}
			\caption{}
			\label{fig:antiSqueezingWithVisibility_dir2}
		\end{subfigure}
	\end{subfigure}
	\caption{(a,b) Probe beam excess noise over shot noise, (c,d) squeezing, and (e,f) anti-squeezing as a function of ${\mathcal{V}_p}^2$ in the probe homodyne detector. We change the visibility by displacing the probe LO along the two directions (direction (1): (a,c,e) and direction (2): (b,d,f)) as shown in Fig.~\ref{fig:beamDisplacement}. To plot the theory curves we use the mean of estimated 4WM gain, and the mean of estimated transmissions on the probe and the conjugate beams across all the data points. Additionally, we use $\varepsilon_{\text{p}}$=0.9 and $\varepsilon_{\text{c}}$=1 (thin solid red curves), and $\varepsilon_{\text{p}}$=$\varepsilon_{\text{c}}$=0 (thick solid orange curves), as defined in the text.}
	\label{fig:model_parameter_validation}
\end{figure}

Experimental measurements showing the variation of the measured probe excess noise with the visibility in the probe homodyne detector can be seen in Fig.~\ref{fig:single_beam_noise_dir1} and ~\ref{fig:single_beam_noise_dir2}. The two graphs represent the measured noise as the visibility is reduced by displacing the LO along the two directions indicated in Fig.~\ref{fig:beamDisplacement}. It can be seen that the noise level stays relatively constant as the visibility is reduced, indicating that the modes surrounding the probe mode have approximately the same noise level. The slight reduction in the measured noise that can be seen in the plots may stem from the non-uniformity of the gain across the output region, such that as the LO is displaced, some of the new thermal modes that are coupled-in are less noisy (experience less gain) than the probe mode. The error bars here and elsewhere in this paper represent one standard deviation statistical uncertainties.

We point out that although there are many ways of changing the visibility, one can always think of the situation by decomposing the LO spatial mode in an orthonormal basis where the probe beam spatial mode serves as an eigenvector. In this representation, any overlap of the LO spatial mode with the eigenvectors orthogonal to the probe beam spatial mode couples in the noise from those orthogonal modes.

\begin{figure}
	\centering
	\includegraphics[width=105mm]{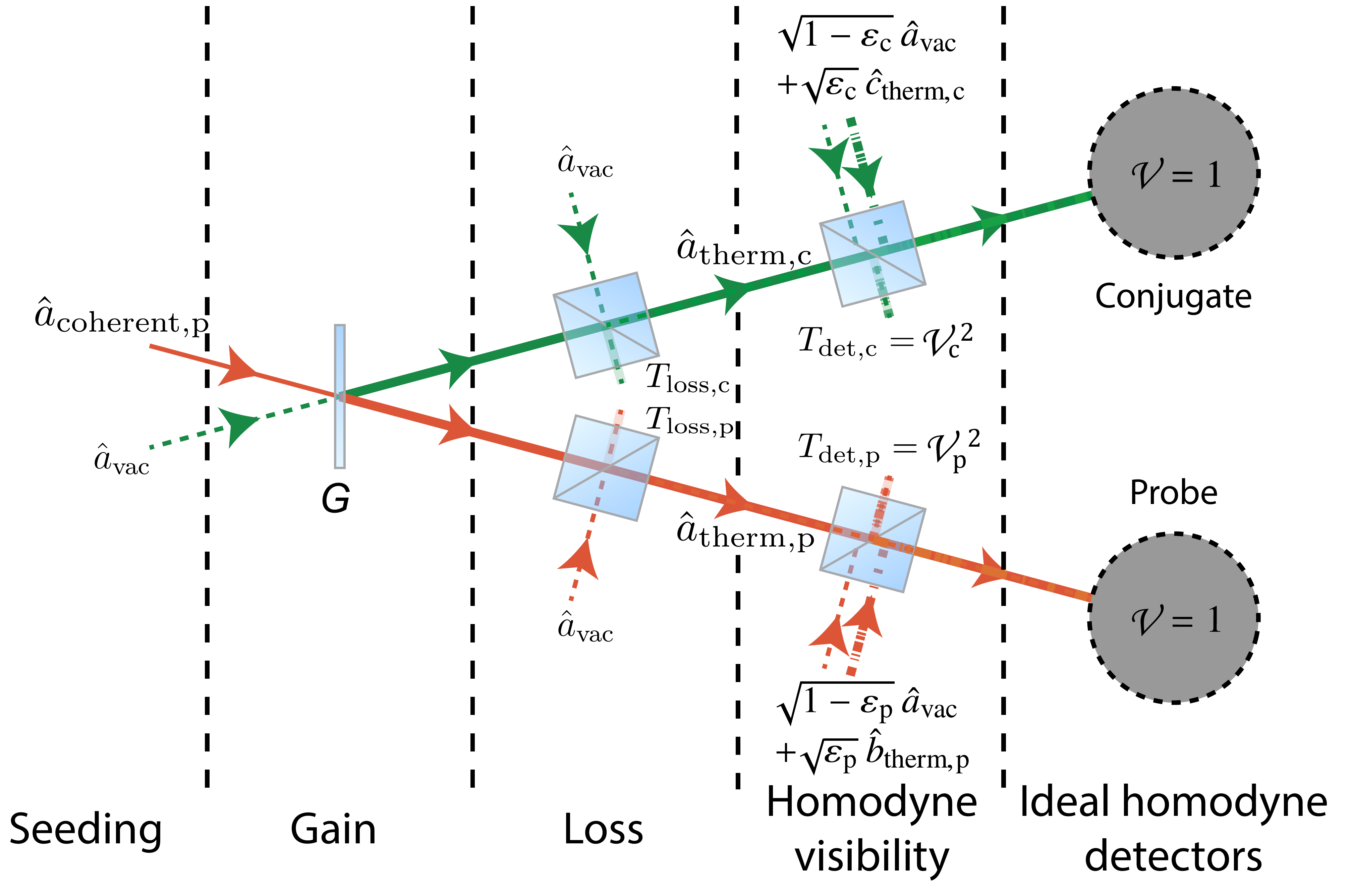}
	\caption{Theoretical model for gain, loss, and homodyne detector visibility in the measurement of quadrature squeezing. $\hat{a}_{\text{coherent,p}}$ represents the annihilation operator for the probe seed, $\hat{a}_{\text{therm,p}}$: probe mode, $\hat{a}_{\text{therm,c}}$: conjugate mode, $\hat{a}_{\text{vac}}$: vacuum mode, $\hat{b}_{\text{therm, p}}$ and $\hat{c}_{\text{therm, c}}$: the additional probe and conjugate thermal modes uncorrelated to the desired modes.    
	The final properties of the beams are modeled by applying the operations to the beam from left to right in the figure. The first element applies gain ($G$) from Rb vapor. We model all the transmission loss in the system with a single beamsplitter having transmittance $T_{loss,p(c)}$ and with vacuum coupling from the second port. We model the effect of an imperfect visibility in each homodyne detector with another beam splitter having transmittance $T_{\text{det,p(c)}}={\mathcal{V}_{\text{p(c)}}}^2$, and with a weighted sum of vacuum and an independent thermal mode with the same noise as the signal mode coupling from the second port. $\varepsilon_{\text{p}}$ and $\varepsilon_{\text{c}}$ represent the fractions of uncorrelated thermal modes entering the second port of the beam splitters.}
	\label{fig:graphical_representation_of_model}
\end{figure}

Having described qualitatively the dependence of the probe beam excess noise on visibility in the probe homodyne detector, we now present a model which explains the phenomenon quantitatively. We use the model to understand the variation of excess probe noise and most importantly of squeezing with the visibilities in the two homodyne detectors. Figure~\ref{fig:graphical_representation_of_model} shows the elements of the model we will use to explain the experimentally observed squeezing levels. In our system, the Rb vapor cell serves as the gain medium. The probe beam frequency lies much closer to a Doppler broadened atomic resonance than does the conjugate beam frequency, and hence the probe suffers noticeably more loss inside the Rb vapor than the conjugate.  To fully model the gain process, one needs to consider the gain and loss distributed over the length of the Rb vapor cell \cite{PhysRevA.78.043816}. In this work, we simplify the model by assuming a single lossless gain medium followed by loss.
During the propagation and the detection process the beams suffer extra losses on the optical elements and the diodes used for detection. We model all these losses using the beam splitters placed in each beam after the lossless gain medium. Another parameter that affects the measurement of noise in our setup is the cumulative electronic noise floor of all the devices used in the detection and the noise generated by any scattered light falling on the diodes. We measure the total noise from the electronic noise floor and the noise due to the scattered light by blocking the probe seed of the LO and the signal beams. We directly subtract this extra noise from all our data, and hence do not put this in the model. 

We model the effect of imperfect visibility in each homodyne detector with additional  beamsplitters. For a single spatial mode system, the loss of visibility can be modeled using a beam splitter with transmittance $T_{det} = \mathcal{V}^2$, where $\mathcal{V}$ is the visibility of the homodyne detector, and with vacuum coming in the second port (Fig.~\ref{fig:graphical_representation_of_model}) \cite{PhysRevA.67.033802,Aoki:06, Polzik1992, Yonezawa1514}. To account for the coupling of other modes in our multi-spatial-mode source, we couple in a combination of vacuum and an uncorrelated thermal mode (having noise equal to that of the signal beams) through the second port of the beam splitter. The fraction of the uncorrelated thermal mode in the second port of the beam splitter is given by $\varepsilon$ with $\varepsilon$= 1 corresponding to all the modes being thermal, and $\varepsilon$= 0 to only vacuum modes (Fig.~\ref{fig:graphical_representation_of_model}). The value of $\varepsilon$ may be different for the two homodyne detectors. We model our system assuming that the spatial modes surrounding the desired signal modes have noise less than or equal to the desired probe (or conjugate) modes. This is a valid assumption for a well aligned system. 

The methodology used to implement our model is described in detail by Anderson et al.\mbox{\cite{PhysRevA.95.063843}}. We find analytical expressions for the quadrature noises of the probe and the conjugate beams as well as the noises of the joint quadratures using the non-commuting algebra package in Ref. \cite{QuntumNotation}
on the model described in Fig.\ref{fig:graphical_representation_of_model}. We show the analytical expressions for the probe beam quadrature noise ($\Delta X^2_{p}$), the conjugate quadrature noise ($\Delta X^2_{c}$), the squeezed joint quadrature noise ($\Delta X^2_{sq}$), and the anti-squeezed joint quadrature noise ($\Delta X^2_{Asq}$) in equations (\ref{probeNoiseEq}-\ref{AntiSqueezedNoiseEq}), where all the noise powers are normalized to a vacuum level of 1: 
 
\begin{equation}
\label{probeNoiseEq}
\Delta X^{2}_{p}=(1+2(G-1)\eta_{p})(\varepsilon_{p}(1-\mathcal{V}_p^2)+\mathcal{V}_p^2)+(1-\varepsilon_{p})(1-\mathcal{V}_p^2)
\end{equation}

\begin{equation}
\label{conjugateNoiseEq}
\Delta X^{2}_{c}=(1+2(G-1)\eta_{c})(\varepsilon_{c}(1-\mathcal{V}_c^2)+\mathcal{V}_c^2)+(1-\varepsilon_{c})(1-\mathcal{V}_c^2)
\end{equation} 

\begin{equation}
\label{squeezedNoiseEq}
\Delta X^{2}_{sq} =1+(G-1)\eta_{p}(\mathcal{V}_{p}^2+\varepsilon_{p}(1-\mathcal{V}_p^2)) +(G-1)\eta_{c}(\mathcal{V}_{c}^2+\varepsilon_{c}(1-\mathcal{V}_c^2))-2\sqrt{G(G-1)}\mathcal{V}_p \mathcal{V}_c \sqrt{\eta_{p} \eta_{c}}
\end{equation}

\begin{equation}
\label{AntiSqueezedNoiseEq}
\Delta X^{2}_{Asq} =1+(G-1)\eta_{p}(\mathcal{V}_{p}^2+\varepsilon_{p}(1-\mathcal{V}_p^2)) +(G-1)\eta_{c}(\mathcal{V}_{c}^2+\varepsilon_{c}(1-\mathcal{V}_c^2))+2\sqrt{G(G-1)}\mathcal{V}_p \mathcal{V}_c\sqrt{ \eta_{p} \eta_{c}}.
\end{equation}
Here $\mathcal{V}_p$ and $\mathcal{V}_c$ are the probe and conjugate detector visibilities, $\varepsilon_{c}$ and $\varepsilon_{p}$ are the fractions of uncorrelated thermal modes coupled into the probe and conjugate detectors, and the $\eta $'s are related to the beam path losses as $\eta_{p} = \sqrt{T_{loss,p}} $ and  $\eta_{c} = \sqrt{T_{loss,c}}$. 

There are other ways to model the noise level of the independent modes surrounding the signal mode. In our model, we are mixing together the required proportions of vacuum modes (which have quite low noise), and thermal modes that have noise equal to the noise of the signal mode (which is determined by the gain of the 4WM process). An equally valid way to adjust the noise level of the surrounding modes would have been to model them as pure thermal modes with gain that could be set independently to the gain experienced by the desired signal beam modes. 
Of course it is possible that both homodyne detectors would have an imperfect visibility with the desired probe and conjugate modes, but in such a way that they are both aligned with some other probe modes, and their associated conjugate modes.  In such a fortuitous case the model here would overstate the effect of imperfect homodyne detector visibility.

The parameters that we use in the model are the 4WM gain, the optical transmissions on the probe and conjugate  beams (in the ``loss'' region), the fraction of the independent thermal modes surrounding the probe and conjugate beams, i.e., $\varepsilon_{\text{p}}$ and $\varepsilon_{\text{c}}$, and the visibilities in the two homodyne detectors. In our system, the losses on the optical elements and diodes (98\% quantum efficient) amount to 12\%-14\%. Additionally, we directly measure the visibility in each homodyne detector. Since the gain and the loss inside the Rb vapor are distributed along the length of the Rb cell, it is difficult to directly measure them accurately. Here we estimate our gain and the transmission losses on the probe and the conjugate using the measured data for the quadrature and joint quadrature noises of the two beams, which we describe below.

\begin{figure}[t]
	\centering
	\begin{subfigure}{0.4\textwidth}
		\centering
		\includegraphics[width=\linewidth]{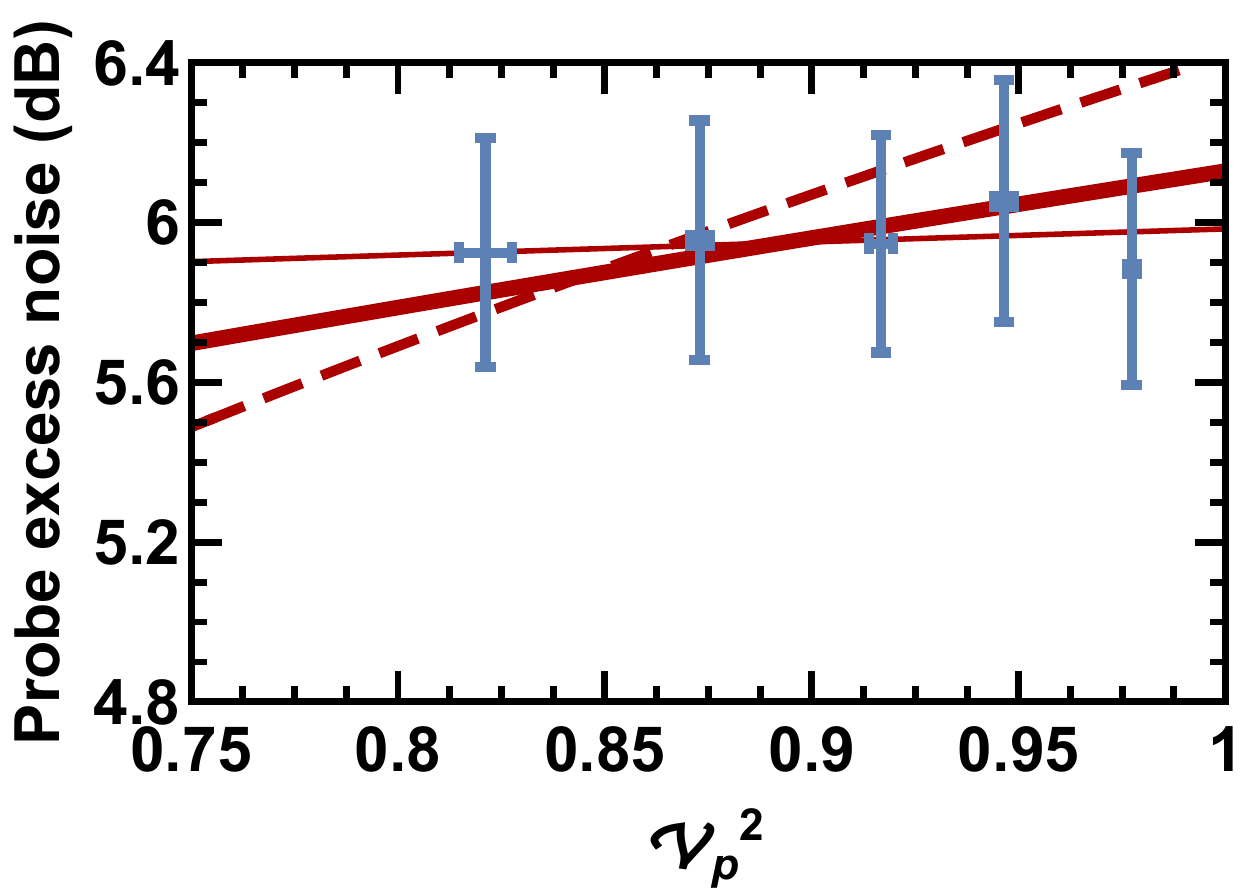}
		\caption{}
	\end{subfigure}
	\hfill
	\begin{subfigure}{0.4\textwidth}
		\centering
		\includegraphics[width=\linewidth]{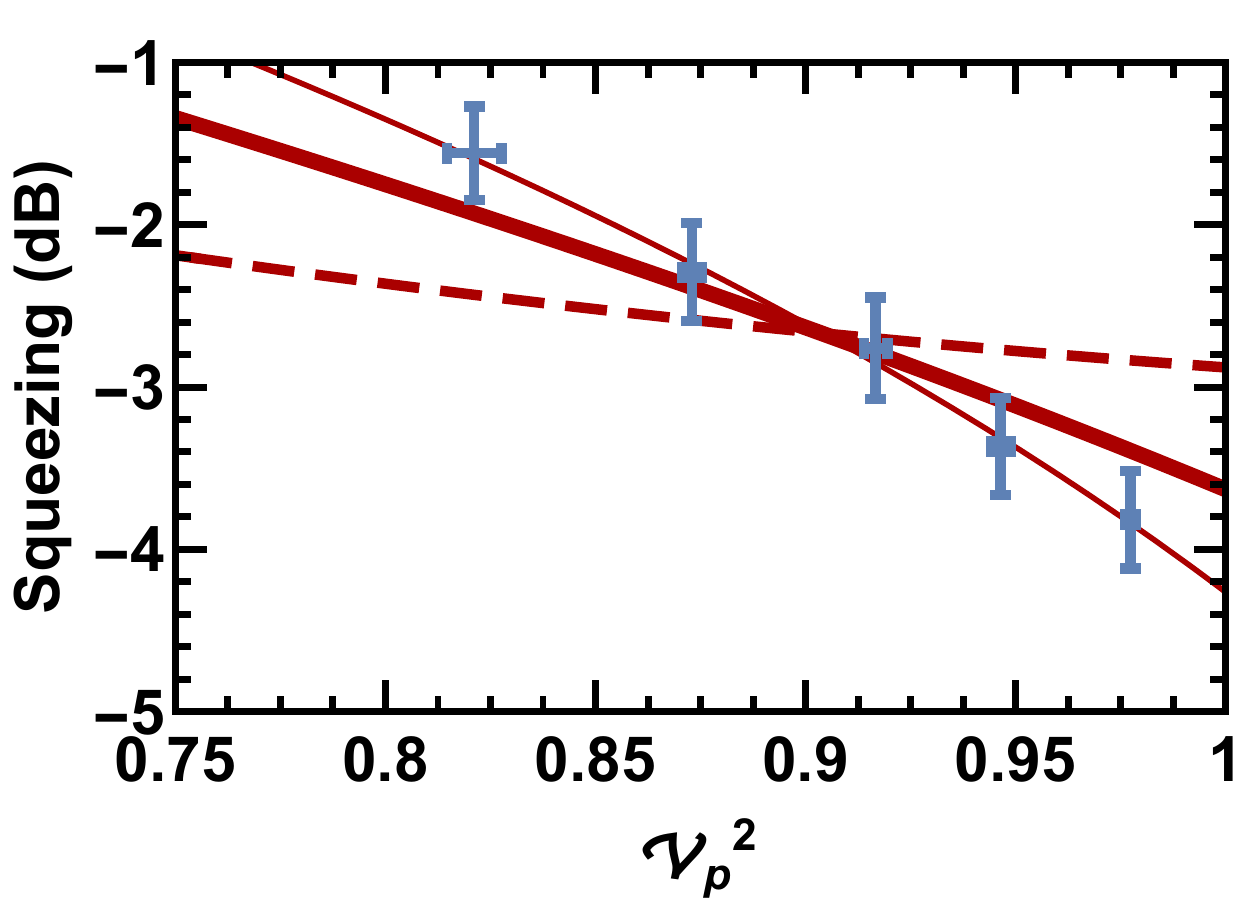}
		\caption{}
		\label{fig:squeezingFitWithDifferentEpsilons}
	\end{subfigure}
	\caption{(a) Probe excesss noise and (b) squeezing versus the square of visibility (${\mathcal{V}_p}^2$) in the probe homodyne detector. Theory plots are computed using the mean of the estimated values of the gain, and the probe and conjugate transmissions across all the data points using $\varepsilon_{\text{c}}$=1 and $\varepsilon_{\text{p}}$=0 (dashed red curve), $\varepsilon_{\text{p}}$=0.5 (thick solid red curve), and $\varepsilon_{\text{p}}$=0.9 (thin solid red curve). The estimated values are $G$=3.9$\pm$0.8, $T_\text{loss,p}$=59\%$\pm$11\%, and $T_\text{loss,c}$=57\%$\pm$13\% (dashed red curve), $G$=3.3$\pm$0.3, $T_\text{loss,p}$=68\%$\pm$5\%, and $T_\text{loss,c}$=69\%$\pm$5\% (thick solid red curve), $G$=3.02$\pm$0.04, $T_\text{loss,p}$=73\%$\pm$1\%,and $T_\text{loss,c}$=77\%$\pm$1\% (thin solid red curve) respectively.}
	\label{fig:noise_fit_with_different_epslons}
\end{figure}

Given homodyne detectors with perfect visibilities, the gain, and the probe and conjugate transmissions in the model could be found from measurements of the probe noise, the conjugate noise, and the squeezing by solving the system of three equations shown in Eqs. (\ref{probeNoiseEq}-\ref{squeezedNoiseEq}) above~\cite{Takeno:07,PhysRevA.75.035802}. With imperfect visibility in the homodyne detectors it is necessary to first assume values for $\varepsilon_{\text{c}}$ and $\varepsilon_{\text{p}}$. Given those values, and the measured homodyne detector visibilities, it is then possible to calculate the gain, and the probe and conjugate transmissions. For the situation where the probe seed is aligned to a region of maximum gain and the output spot is smaller than the gain annulus, it is reasonable to assume that $\varepsilon$ should be near 1. For our analysis here we use values $\varepsilon_{\text{p}}$=0.9 and $\varepsilon_{\text{c}}$=1. 
These choices result in the most consistent values for the loss and gains near the measured values.

To see this we estimated the value of the gain parameters and the probe and the conjugate transmissions using other values for $\varepsilon_{\text{p}}$ and $\varepsilon_{\text{c}}$. While holding $\varepsilon_{\text{c}}$ =1, we perform the same analysis using the values of $\varepsilon_{\text{p}}$=0.5 and $\varepsilon_{\text{p}}$=0 that we did above with $\varepsilon_{\text{p}}$=0.9, and plot the theoretical estimates using these values of $\varepsilon_{\text{p}}$ in Fig.~\ref{fig:noise_fit_with_different_epslons}. We compare these results against the experimental data taken by changing the probe homodyne detector visibility by moving the probe LO along the direction (1). The $\varepsilon_{\text{p}}$=0.5 and $\varepsilon_{\text{p}}$=0 curves in Fig.~\ref{fig:squeezingFitWithDifferentEpsilons} do not follow the experimental data well, and extrapolate to squeezing levels at perfect visibility lower than that experimentally observed at the highest visibility point. Not evident in the plots is that choosing $\varepsilon_{\text{p}}$ significantly different from 0.9 gives larger variation in the estimated gain and transmission parameters across different data points taken with the same source conditions, and only with different probe homodyne visibilities. For instance, choosing values of $\varepsilon_{\text{c}}$=1 and $\varepsilon_{\text{p}}$=0, the estimated values turn out to be 3.9$\pm$0.8 for gain, and 59\%$\pm$11\% and 57\%$\pm$13\% for probe and conjugate transmissions respectively. Since all the data is taken with nominally the same source conditions, the estimated parameters for different data points should agree closely with each other. Hence, higher variation in estimated parameters across the data as $\varepsilon_{\text{p}}$ goes to 0 indicates that the correct value of $\varepsilon_{\text{p}}$ is close to 1. We performed the same analysis for the data taken by changing the visibility along the direction (2) and observed a similar behavior.

In Fig.~\ref{fig:model_parameter_validation}, we show the probe beam excess noise, the squeezing, and the anti-squeezing as we change the visibility in the probe homodyne detector. We kept the source conditions (the gain and the probe and conjugate transmissions) constant and changed only the probe homodyne detector visibility while taking data. We use the above-mentioned values of the fractions $\varepsilon_{\text{p}}$ and $\varepsilon_{\text{c}}$ to estimate the gain and the probe and conjugate transmissions by solving the system of equations at each point. At each point, our estimated value of the gain matches closely with the gain we measured with a DC power meter. (In the experiment, we measure the gain by using the relationship,$P_{\text{conj,DC}}$ = ($G$-1)*$P_{\text{probe seed,DC}}$, where $P_{\text{conj,DC}}$ and $P_{\text{probe  seed,DC}}$ are the DC powers of the conjugate beam and the probe seed beam, respectively). The estimated losses are a little higher than the optical transmission loss outside the cell mentioned above. This is expected as the simultaneous action of gain and loss inside the Rb vapor cell places extra losses on the beams.

To obtain a theory curve on each of the plots in Fig.~\ref{fig:model_parameter_validation}, we take the estimated values of the gain and the optical transmissions at each point and find a mean value of gain and the optical transmissions across all the data points. Our estimated values are 3.02$\pm$0.04 for gain, 73\%$\pm$1\% for the probe transmission, and 77\%$\pm$1\% for the conjugate transmission for direction (1). The estimated values for direction (2) are 3.02$\pm$0.06 for gain, 72\%$\pm$2\% for the probe transmission, and 78\%$\pm$2\% for the conjugate transmission. Since the data was taken with nominally constant source conditions we use the average values of these parameters in addition to the assumed values of $\varepsilon_{\text{p}}$ and $\varepsilon_{\text{c}}$. We find that the theory curve agrees well with the data in all the plots. The uncertainty here and elsewhere in the parameter estimation represents one statistical standard deviation calculated using the estimated parameter values for all the data points.

It is instructive to compare the results for our multi-spatial-mode system to one producing the same level of squeezing in a single spatial mode for both the probe and conjugate, i.e., we want to compare single-spatial-mode and multi-spatial-mode systems that would have the same measured squeezing level for perfect visibilities in the two homodyne detectors ($\mathcal{V}_i =1$).To that end we plot in Fig.~\ref{fig:model_parameter_validation} the thick solid orange curves calculated using the same gain and transmission values as for the red curves, but with $\varepsilon_{\text{p}}$ = $\varepsilon_{\text{c}}$ =0. We see that the behavior of the single-spatial mode system differs significantly from our data for the probe beam excess noise (Figs.~\ref{fig:single_beam_noise_dir1} and~\ref{fig:single_beam_noise_dir2}), and especially for the measured squeezing (Figs.~\ref{fig:squeezingWithVisibility_dir1} and~\ref{fig:squeezingWithVisibility_dir2}). 

\begin{figure}[t]
	\centering
	\includegraphics[width=0.5\textwidth]{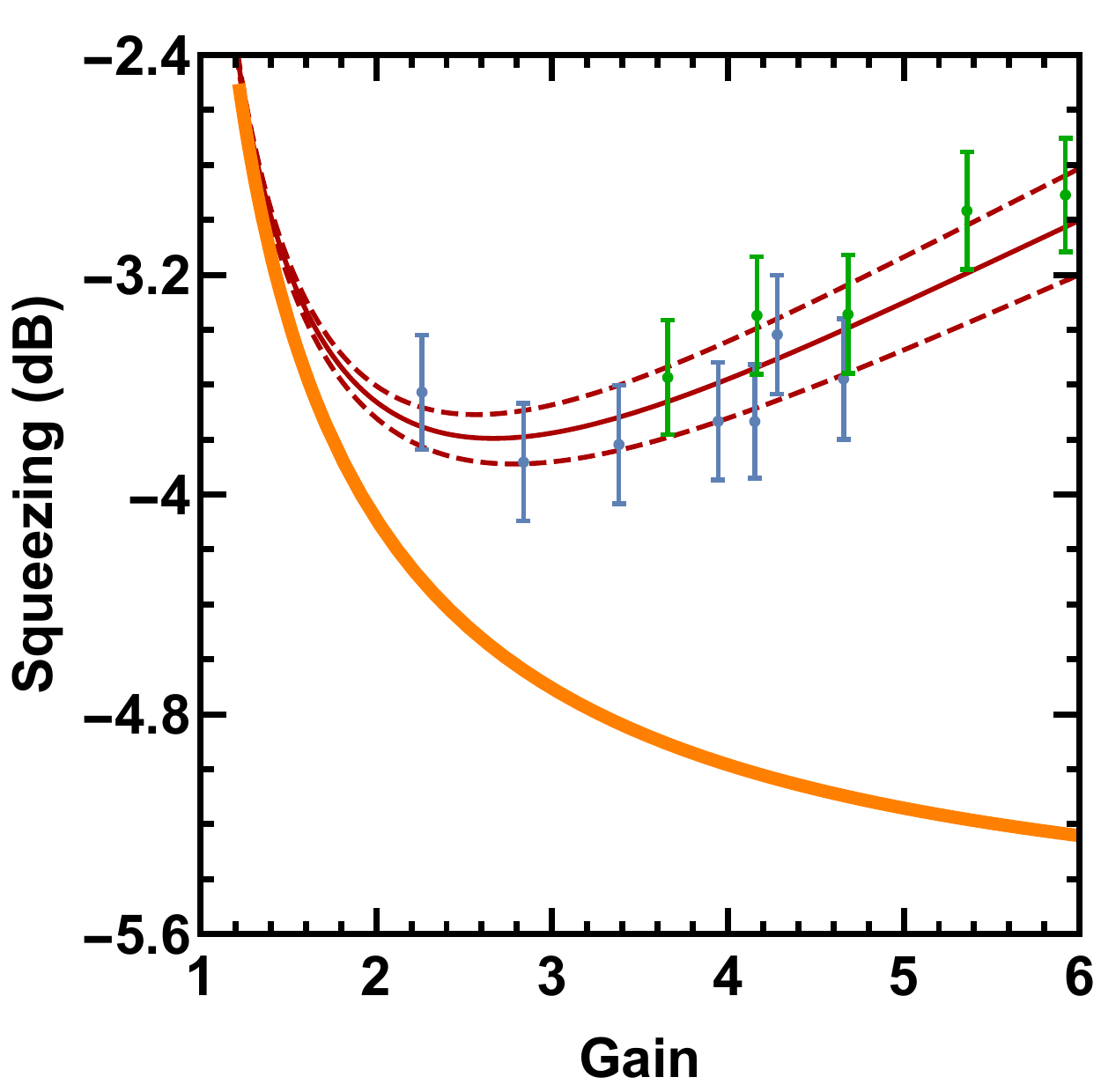}
	\caption{Measured squeezing levels as a function of 4WM gain. For each of these data points, the mean visibilities in the probe and the conjugate homodyne detectors lie within 0.986$\pm$0.001 and 0.986$\pm$0.002. The estimated values of probe and the conjugate transmissions for the data points are 74\%$\pm$2\% and 78\%$\pm$2\% respectively. The green data points were taken at a higher cell temperature, and thus higher rubidium vapor density. The green points show slightly lower transmission than the blue points but the difference is not significant for the present analysis. We use the mean values of the estimated transmissions to calculate the theory curves. Additionally, we use $\varepsilon_{\text{p}}$=0.9 and $\varepsilon_{\text{c}}$=1 in plotting the solid and dashed red curves. The thin solid red curve utilizes the mean visibilities in each homodyne detector and we plot the two thin red dashed curves with both the visibilities one standard deviation away from the mean values. The x-axis in the plot represents the estimated 4WM gain from our model. To demonstrate the effect of visibility in a single-spatial mode system, we plot the thick solid orange curve using the mean values of the estimated transmissions, the mean homodyne detector visibilities, and by using $\varepsilon_{\text{p}}$=$\varepsilon_{\text{c}}$=0.}
	\label{fig:gain_squeezing}
\end{figure}

\section{Variation in measured squeezing with system gain}
Having validated our model in a case where we kept the source operation the same and intentionally varied the homodyne detector visibility, we now turn to the question of how the measured squeezing level varies as the gain of the system is changed, holding the homodyne detector visibility near its maximum value. To study the behavior, we change the gain by using a combination of varying the one photon detuning of the pump beam and by changing the density of Rb atoms inside the vapor cell by changing the temperature\cite{PhysRevA.78.043816}. We keep the visibilities of the probe and the conjugate homodyne detectors fixed at 0.986$\pm$0.001 and 0.986$\pm$0.002 respectively. For each gain value, we measure the squeezing, anti-squeezing, and the probe and conjugate beam excess noises (Fig.~\ref{fig:gain_squeezing}). As noted earlier, the probe beam suffers from lensing due to the cross-Kerr modulation induced by the pump beam. Changing the pump detuning changes this lensing. The conjugate beam is further detuned from the atomic resonance, and thus is less strongly affected by the presence of the pump beam. The lensing effect causes the homodyne detector visibility to change with pump laser tuning. We were able to hold the detector visibilities at the value quoted above by making small realignments of the beam pointing. 

To understand the behavior of the squeezing as a function of gain shown in Fig.~\ref{fig:gain_squeezing}, we take $\varepsilon_{\text{c}}$=1, $\varepsilon_{\text{p}}$=0.9, and then, as before, solve for the gain, and the probe and conjugate transmissions. Since we took data without significantly changing the optical losses, we estimated similar values of probe and conjugate transmissions for all the data points, which are $74\%\pm2\%$ and $78\%\pm2\%$ respectively. The red solid curve gives the theoretical prediction from our model taking into account the surrounding uncorrelated thermal modes. We use the mean of the estimated probe and the conjugate transmissions across all the data points to calculate the theory curves. Using our model, we also generate dashed red curves with both homodyne detector visibilities one standard deviation higher and one standard deviation lower. In agreement with the data, the model predicts that the measured squeezing level decreases with gain once the gain reaches a certain value.
The thick solid orange curve shows the expected squeezing if $\varepsilon_{\text{c}}$=$\varepsilon_{\text{p}}$=0, i.e., if the squeezed probe and the conjugate modes were surrounded by vacuum rather than noisy thermal modes. In contrast to the data and multi-spatial-mode theory, in this case the measured squeezing would increase monotonically with gain.

\begin{figure}[t]
	\centering
	\begin{subfigure}{0.4\textwidth}
		\centering
		\includegraphics[width=\linewidth]{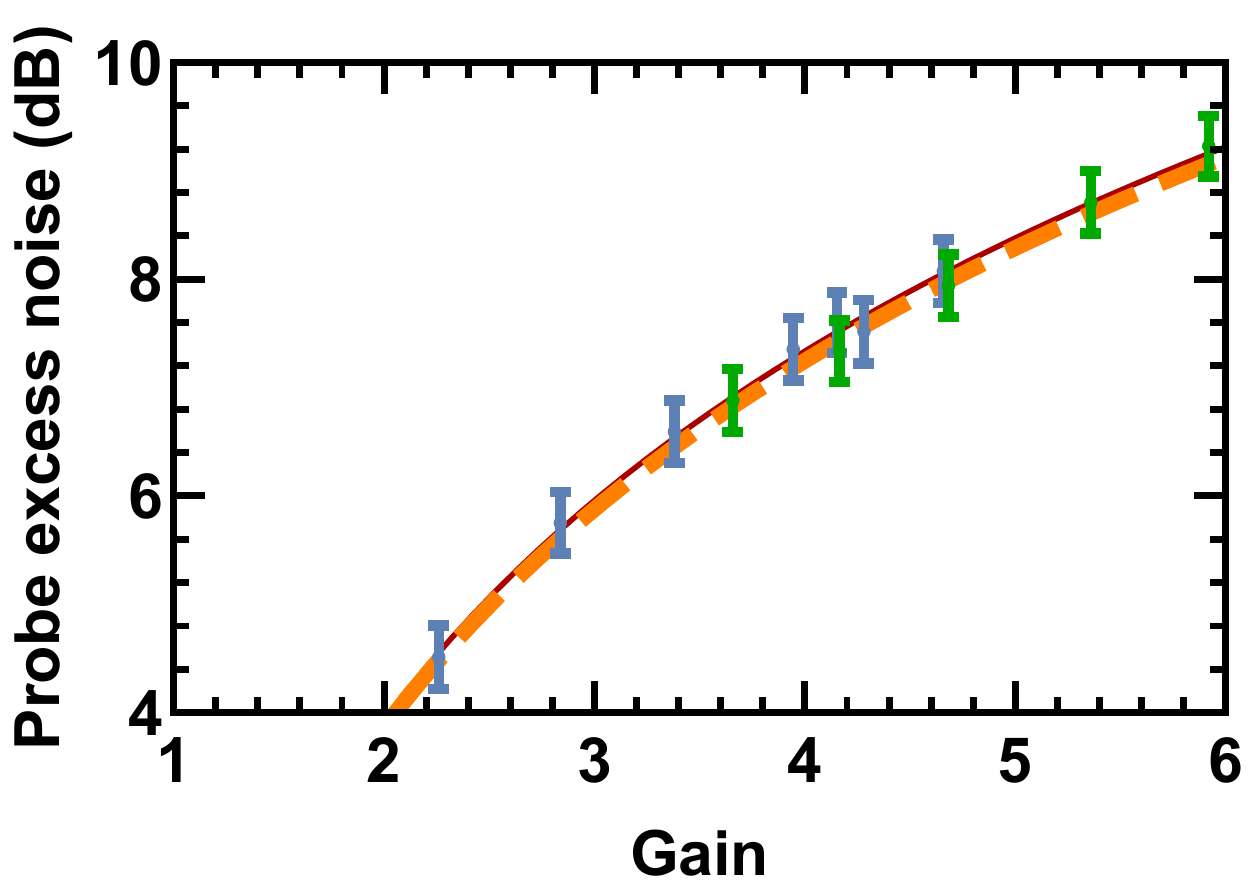}
		\caption{Probe excess noise.}
		\label{fig:squeezingWithVisibilityCH1}
	\end{subfigure}
	\hfill
	\begin{subfigure}{0.4\textwidth}
		\centering
		\includegraphics[width=\linewidth]{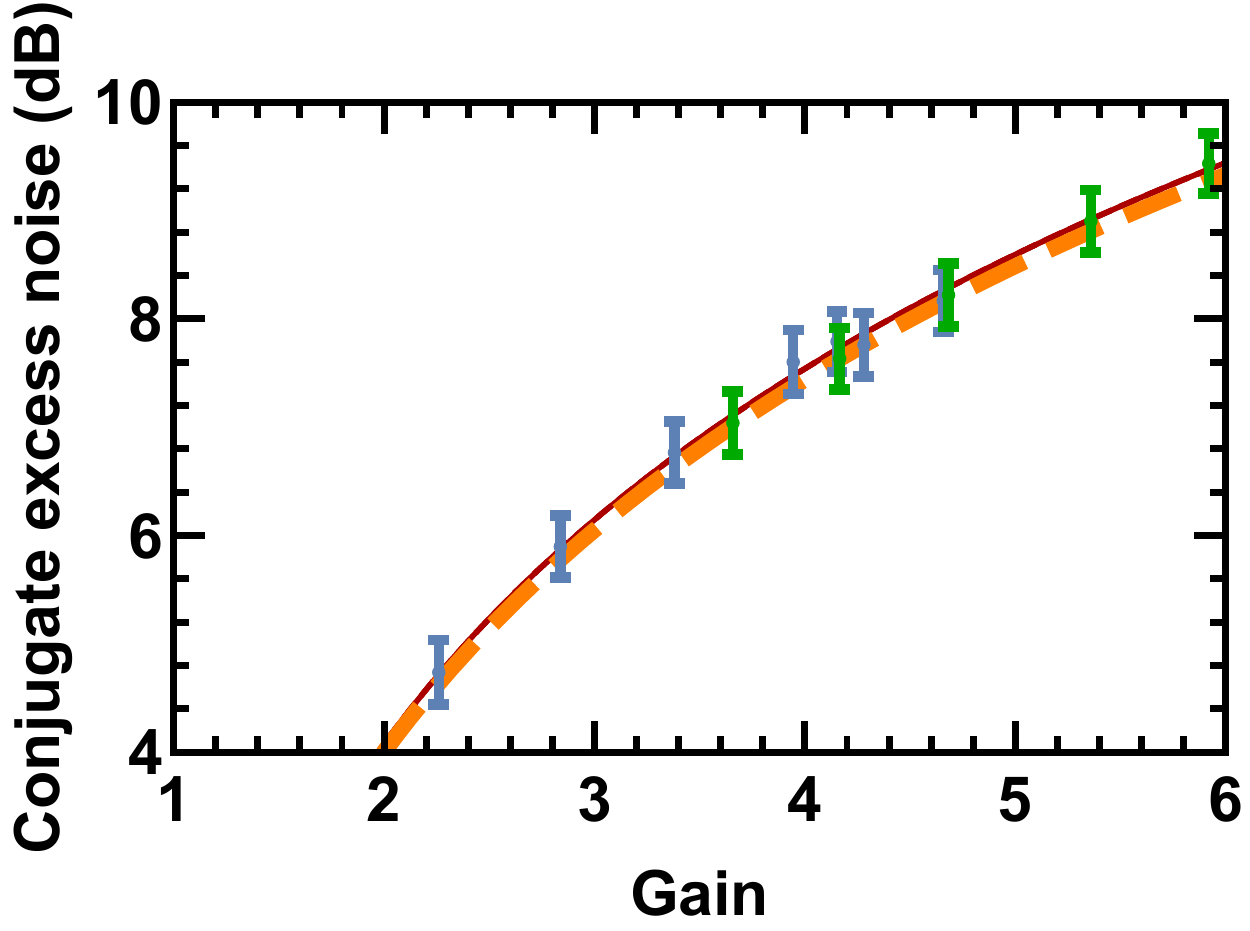}
		\caption{Conjugate excess noise.}
		\label{fig:squeezingWithVisibilityCH2}
	\end{subfigure}
	\hfill
	\begin{subfigure}{0.4\textwidth}
		\centering
		\includegraphics[width=\linewidth]{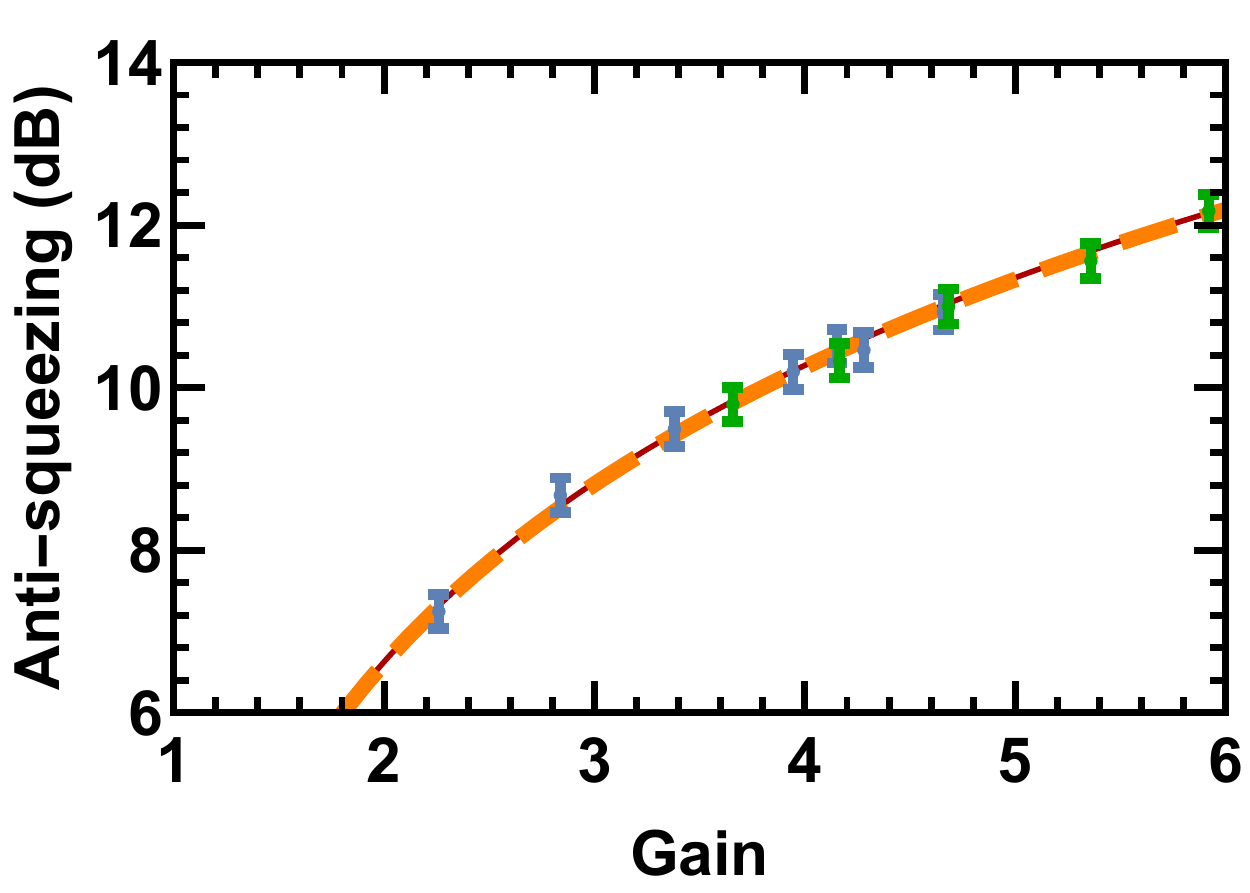}
		\caption{Measured antisqueezing.}
		\label{fig:squeezingWithVisibilityCH3}
	\end{subfigure}
	\caption{Variation of (a) probe excess noise, (b) conjugate excess noise, and (c) anti-squeezing as a function of gain. The parameters used here to generate the theory curves are the same used in generating the theory curves in Fig.~\ref{fig:gain_squeezing}. The solid red curves assume $\varepsilon_{\text{p}}$=0.9 and $\varepsilon_{\text{c}}$=1 and the orange curves use $\varepsilon_{\text{p}}$=$\varepsilon_{\text{c}}$=0, with the probe and conjugate transmission values constant for both the curves.}
	\label{fig:model_parameter_validation_varying_gain}
\end{figure}

We also plot the variation of the probe and conjugate excess noises and the anti-squeezing as a function of 4WM gain in our system (Fig.~\ref{fig:model_parameter_validation_varying_gain}). Since we took all these data points at high visibilities in both the homodyne detectors, the difference between the thermal mode coupling (thin solid red curves) and the vacuum mode coupling (thick dashed orange curves) due to the loss of the visibility in homodyne detectors is not very significant for these quantities (in contrast to the squeezing). The purpose of these plots is to check that the model, with the assumption of constant probe and conjugate transmissions, reproduces all of the observations, which it does.

\begin{figure}
	\centering
	\includegraphics[width=0.5\linewidth]{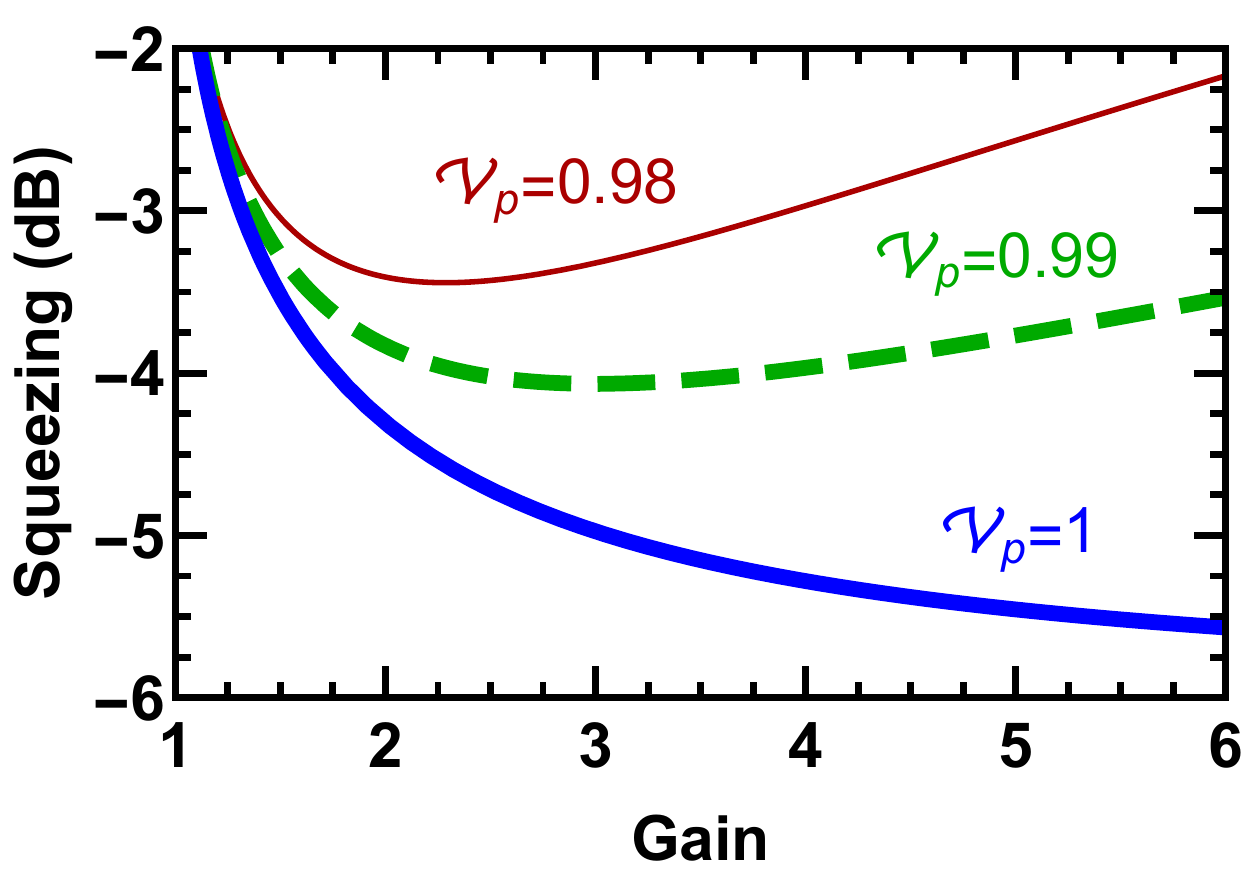}
	\caption{Theoretical squeezing versus gain at different probe and conjugate homodyne detector visibilities assuming only the thermal mode coupling due to the imperfect visibility. The probe and the conjugate homodyne visibilities used in the three curves are 98\% (thin solid red), 99\% (dashed green), and 100\% (thick solid blue). Additionally, we use our estimated values of 74\% probe transmission and 78\% conjugate transmission in calculating the theory curves.}
	\label{fig:TheorecticalSqueezingWithGainVisibility}
\end{figure}

In most OPO systems for producing squeezed light the measurement of the squeezing level is limited by the phase noise between the LO and the squeezed state in a homodyne detector, which in the literature ranges from $0.1\textsuperscript{o}$ to $3.9\textsuperscript{o}$ \cite{PhysRevLett.117.110801,Takeno:07, doi:10.1063/1.2335806}. In OPO systems, single-mode squeezing levels of 9 dB and 15 dB below the vacuum level have been reported by reducing the phase noises to $1.5\textsuperscript{o}$ and $0.1\textsuperscript{o}$ respectively \cite{Takeno:07, PhysRevLett.117.110801}. In our system, we have phase noises (rms values) of $0.45\textsuperscript{o}$ and $0.55\textsuperscript{o}$ in the probe and the conjugate homodyne detectors, respectively. The total phase noise is $\approx\!1\textsuperscript{o}$, which is sufficiently small that it is not the dominant factor in limiting the measured squeezing in our experiment. Instead, in the current state of the experiment, the imperfect visibilities in the two homodyne detectors are the dominant factors in limiting the measured squeezing.
Figure~\ref{fig:TheorecticalSqueezingWithGainVisibility} shows the behavior for different visibilities in the two homodyne detectors.  Routinely in the lab we get 98\% visibility.  As shown earlier we can, when paying close attention to the visibility, get 98.6\%.  This limits our measured squeezing to about 4 dB below the shot noise at a gain of about 3, even though the source is capable of achieving a higher gain and a higher level of squeezing. If, in the future, we can shape the probe and conjugate LO beams to better match the signal beams we should be able to push the visibility further and improve the measured squeezing level significantly as shown in Fig.~\ref{fig:TheorecticalSqueezingWithGainVisibility}.

\section{Conclusion}
We have demonstrated the effect of homodyne detector visibility on squeezing measurements in our multi-spatial-mode two-mode squeezed state system. Given imperfect visibilities in the two homodyne detectors and constant transmission losses on the two beams, the measured squeezing maximizes for a certain gain value and then starts to decrease with increasing gain. The coupling of excess noise from uncorrelated thermal modes due to the imperfect visibilities in the two homodyne detectors reduces the measured squeezing much faster than in a single-spatial-mode squeezed state system where the imperfect overlap in the homodyne detector couples only to vacuum modes.

\section*{Funding}
National Science Foundation (NSF) Grant 1708036 and Air Force Office of Scientific Research (AFOSR) Grant FA9550-16-1- 0423.

\section*{Disclosures}
The authors declare that there are no conflicts of interest related to this article.


\end{document}